\documentclass{elsart5p}
\usepackage{amssymb}
\usepackage{graphicx}  
\usepackage{wasysym}

\begin{document}
\begin{frontmatter}

\title{High-precision efficiency calibration of a high-purity co-axial germanium detector}
\author[label1]{B. Blank}\footnotemark[1]
\author[label1]{J. Souin} 
\author[label1]{P. Ascher} 
\author[label1]{L. Audirac} 
\author[label1]{G. Canchel}  
\author[label1]{M. Gerbaux} 
\author[label1]{S. Gr\'evy} 
\author[label1]{J. Giovinazzo} 
\author[label1]{H. Gu\'erin} 
\author[label1]{T. Kurtukian Nieto} 
\author[label1]{I. Matea}\footnotemark[2]
\author[label2]{H. Bouzomita} 
\author[label2]{P. Delahaye} 
\author[label2]{G.F. Grinyer} 
\author[label2]{J.C. Thomas}

\address[label1]{Centre d'Etudes Nucl\'eaires de Bordeaux Gradignan - 
        UMR 5797, CNRS/IN2P3 - Universit\'e de Bordeaux, Chemin du Solarium, BP 120,
        33175 Gradignan Cedex, France }
\address[label2]{Grand Acc\'el\'erateur National d'Ions Lourds, 
        CEA/DSM - CNRS/IN2P3, Bvd Henri Becquerel, BP 55027, F-14076 CAEN Cedex 5, France}
        
\begin{abstract}

A high-purity co-axial germanium detector has been calibrated in efficiency to a precision of about 0.15\% over 
a wide energy range. High-precision scans of the detector crystal and $\gamma$-ray source measurements have 
been compared to Monte-Carlo simulations to adjust the dimensions of a detector model. For this purpose,
standard calibration sources and short-lived online sources have been used. The resulting efficiency calibration
reaches the precision needed e.g. for branching ratio measurements of super-allowed $\beta$ decays for tests of
the weak-interaction standard model.

\end{abstract}

\begin{keyword}
gamma-ray spectroscopy \sep super-allowed beta transitions \sep germanium detector \sep Monte-Carlo simulations

\PACS 
07.85.-m \sep 29.30.Kv \sep 23.20.-g
\end{keyword}
\end{frontmatter}


\footnotetext[1]{Corresponding author: B. Blank, blank@cenbg.in2p3.fr}
\footnotetext[2]{Permanent address: Institut de Physique Nucl\'eaire, 15 rue Georges Cl\'emenceau, 91406 Orsay Cedex, France}

\setcounter{footnote}{0}

\section{Introduction}

Many spectroscopic studies in nuclear physics require only modest precisions (typically of the order of 1-10\%) simply
because nuclear-structure or astrophysical models are limited in their precision and therefore in their predictive power
due to the lack of a high-precision standard model in nuclear physics. However, studies at the interface between nuclear
and particle physics, where the nuclear $\beta$ decay is used as a probe, require precisions which go well beyond the
above mentioned level. Nuclear 0$^+ \rightarrow$ 0$^+$ $\beta$ decay is presently the most precise means to determine the weak-interaction
vector coupling constant $g_V$ which, together with the coupling constant for muon decay, allows the determination of the
$V_{ud}$ matrix element of the Cabibbo-Kobayashi-Maskawa (CKM) quark mixing matrix. To determine this matrix element, 
the $\beta$-decay Q value, the half-life and the super-allowed 0$^+ \rightarrow $ 0$^+$ branching ratio have to be measured with
a relative precision of about 0.1\%. Q-value determination has become "quite easy" with the use of Penning-trap mass 
spectrometry where precisions well below 1~keV, equivalent to some 10$^{-4}$ relative precision, are now routinely 
reached for most of the nuclei of interest. The half-life is usually determined by $\beta$-particle or $\gamma$-ray 
counting and precisions of the half-lives well below 10$^{-3}$ are obtained~\cite{hardy09}.

In these measurements, the precise determination of the branching ratio remains the tricky part, in particular in more
exotic nuclei (e.g. nuclei with an isospin projection T$_z$= -1) where the non-analogue branches, i.e. the branches 
other than the 0$^+ \rightarrow$ 0$^+$ $\beta$-decay branch, are of the same order of magnitude as the analogue branch. As, due to
a continuous spectrum, it is 
extremely difficult to determine these branching ratios by a measurement of the $\beta$ particles, the branching ratios
are usually determined by detecting $\gamma$ rays de-exciting the levels populated by $\beta$ decay by means of germanium 
detectors. Therefore, in order to determine a branching ratio with a precision of the order of 0.1\%, one needs to know
the absolute efficiency of a germanium detector with a similar or better precision.

To our knowledge, there is presently one germanium detector which is efficiency calibrated to such a 
precision~\cite{hardy02,helmer03,helmer04}. The calibration of a single-crystal high-purity co-axial germanium detector
we present here will therefore follow to some extent the procedure used by Hardy and co-workers.

The aim of the present work is to construct a detector model using a simulation tool able to calculate detection 
efficiencies in different environments for any radioactive source or $\gamma$-ray energy. For this purpose, we have used a 3D detector scan 
and 23 different radioactive sources to calibrate our detector in full-energy peak as well as in total efficiency and we have tuned our 
detector model to match the source measurements. As will be laid out in detail below, we used 10 sources to determine 
the total-to-peak (T/P) ratio, and 14 sources for the peak efficiency determination, a dedicated $^{60}$Co source being used 
in both series of measurements. For the simulations, we used mostly the CYLTRAN code~\cite{halbleib86} which was 
upgraded to allow for the simulation of complete decay schemes of radioactive sources. We also added the process
of positron annihilation-in-flight originally not present in the code. In a later stage of our work, we implemented 
our detector model also in GEANT4~\cite{geant4} and found perfect agreement between the two codes, with the 
CYLTRAN code being much faster than GEANT4 and GEANT4 allowing for more flexibility of the geometry of the 
problem simulated.

\section{Detector, electronics, and data acquisition}

The detector we purchased for the present work is an n-type high-purity co-axial germanium detector with a relative efficiency
of about 70\%. An n-type detector is important in particular for detecting low-energy $\gamma$ rays, as the thick dead zone 
for an n-type detector is on the inner surfaces of the detector and, in particular, not on the front surface facing the 
radioactive sources. We have chosen an aluminum entrance window instead of a much more fragile beryllium window, because
the detector will travel to different laboratories. A large dewar ensures an autonomy of the detector of close to four days.

\begin{figure}[hht]
\begin{center}
\includegraphics[width=0.45\textwidth,angle=-0.4]{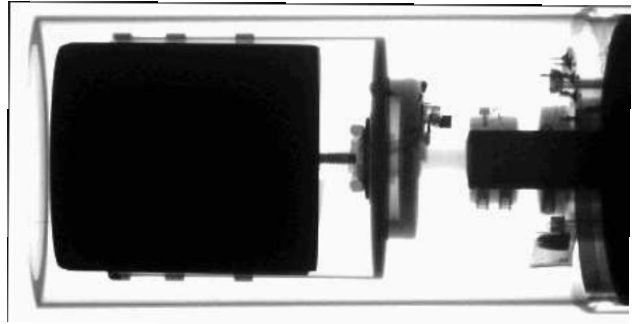}
\caption{X-ray photograph of the germanium detector. The slight tilt of the crystal with respect to the detector housing 
         (about 1 degree) is visible.
         } 
\label{fig:xray}
\end{center}
\end{figure}

A X-ray photography of the detector (Fig.~\ref{fig:xray}) shows a slight tilt of approximately 1$^o$ of the detector crystal 
in the aluminum can. GEANT4 simulations lead us to the conclusion that this tilt has no influence on the results of the 
present work. The initial manufacturer characteristics of the detector, used as a starting point for the simulations, 
are given in Tab.~\ref{tab:det}.

As in the work of Hardy and co-workers, we used a fixed source - detector distance of 15~cm. This distance allows us to
reach a positioning precision of 10$^{-3}$ even in more difficult online conditions with radioactive sources deposited on a tape 
transport system.

The experiment electronics consists mainly of an ORTEC 572A spectroscopy amplifier, an ORTEC 471 timing filter amplifier, 
and an ORTEC 473A discriminator. The signal from the discriminator is then sent to a LeCroy 222
gate generator (gate length typically a few micro-seconds) which then triggers the data acquisition (DAQ). The measurement duration
is determined with a high-precision pulse generator (relative precision of 10$^{-5}$) fed into a CAEN VME scaler (V830) in the DAQ.

The data acquisition is a standard GANIL data acquisition~\cite{ganildaq} using VME modules. In addition to the scaler, we use a V785 ADC 
from CAEN for the energy signals and the TGV trigger unit built by LPC Caen~\cite{tgv}.

\begin{table}[hht]
\begin{center}
\caption{Detector characteristics used in the modelling of the germanium detector. The vendor specifications
         are compared to the finally adopted characteristics.
         } 
\begin{tabular}{lrr}
&& \\
\hline \rule{0pt}{1.3em}
                        & vendor & adjusted  \\
                        &  specifications & ~~~~parameters \\
[0.5em] \hline \rule{0pt}{1.3em}
length of crystal       & 79.2 mm               & 78.10 mm \\
radius of crystal       & 34.8 mm               & 34.66 mm \\
length of central hole  & 59.5 mm\footnotemark & 71.59 mm \\
radius of central hole  & 5.0 mm                & 7.8 mm \\
external dead zone      & $<$ 0.5 $\mu$m          & 0.5 $\mu$m \\
internal dead zone      & 0.5 mm                & 0.5 mm \\
back-side dead zone     & 0.5 mm                & 1 mm \\
distance window - crystal & 5 mm                & 5.73 mm \\
entrance window thickness~~ & 7 mm                & 7 mm \\   
[0.5em] \hline
\end{tabular}
\label{tab:det}
\end{center}
\end{table}

\footnotetext[1]{A value given later by the vendor was 70.0(5) mm.}

The dead-time correction is performed by means of a 1 kHz pulser sent directly to the scaler and passed through a veto module
(Phillips 758), where the veto is generated by the BUSY signal of the DAQ. The live-time (LT) is the ratio between the second 
and the first scaler values and the dead-time (DT) is equal to 1 - LT. To test the dead-time correction, a radioactive source 
($^{137}$Cs) was mounted on the source holder at 15~cm from the detector entrance window. A measurement yielded a first result 
for the counting rate in the detector. We then added other triggers from a pulse generator. Thus the trigger rate of the DAQ 
was steadily increased without affecting the number of $\gamma$ rays emitted from the $^{137}$Cs source hitting the detector. 
Without dead-time correction, the apparent counting rate from the source decreases with increasing total trigger rate. When 
corrected for the acquisition dead time as described above, we recover the source counting rate without dead-time. We 
performed similar tests also with a $^{60}$Co source (twice higher $\gamma$-ray energies) and found equivalent results.

\begin{figure}[hht]
\begin{center}
\includegraphics[width=0.26\textwidth,angle=-90]{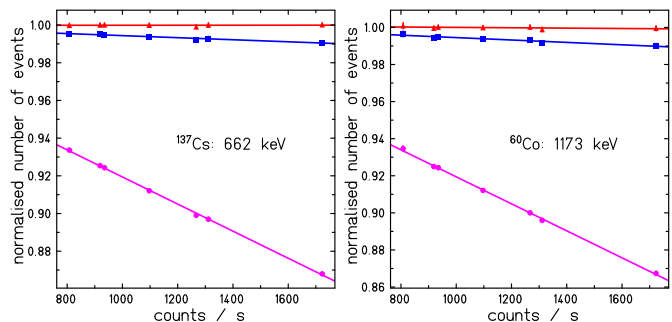}
\caption{Dead-time and pile-up corrections as described in the text. The constant counting rate of a $^{137}$Cs source 
         (left-hand side) is recovered after correcting for the acquisition dead-time and the signal pile-up. Similar results
         are obtained for the $^{60}$Co source (right-hand side). The purple curves (full circles) correspond to the uncorrected data, the blue
	     lines (full squares) to the data corrected only for the dead-time, and the red lines (full triangles) represent the data corrected for both, 
         the dead-time and the pile-up.
         } 
\label{fig:dead_pile}
\end{center}
\end{figure}

Another concern when aiming for very high detection efficiency precision is the pile-up of radiation in the detector. Due to summing 
of signals from different events, counts are removed from the full-energy peak of a $\gamma$ ray and moved to higher energy.
We correct this by assuming a Poisson distribution of the events around a measured average count rate. With this assumption, 
we can determine the pile-up probability once we have defined a "pile-up time window"~\cite{grinyer07}. This time
window was determined in a similar fashion as the dead-time correction. The full-energy peak counting rate was determined
for a fixed source ($^{137}$Cs or $^{60}$Co). In a second step, a low-energy source ($^{57}$Co) was approached more and more 
to increase the trigger rate, but also the pile-up. In the analysis, the pile-up time window was varied to achieve a full-energy 
peak rate of the fixed source independent from the total counting rate of the detector. The results of this procedure are shown in 
Fig.~\ref{fig:dead_pile}. It was found experimentally that the pile-up time window depends, as expected, linearly on the shaping time of the 
amplifier (2.75 $\times$ shaping time) and is, at least in the limit of the precision we achieved, independent 
of the $\gamma$-ray energy. This last finding is not necessarily expected, as, e.g. in a too large acquisition window, a larger
signal coming after a smaller one will "erase" this smaller signal in our peak-sensing ADC.

All measurements except those for the dead-time and pile-up correction measurements, in particular those for the determination 
of the absolute efficiency with the $^{60}$Co source, were performed with counting rates well below 1000 counts per second 
where the dead-time and the pile-up are relatively small and can be therefore reasonably well corrected for.

\section{Monte-Carlo simulations}

As explained above, the aim of this work is to construct a model for a simulation code which allows us to determine the detection
efficiency of the germanium detector in different environments, for any source and $\gamma$-ray energy. Several codes have been 
used for similar purposes. We have chosen the CYLTRAN~\cite{halbleib86} and the GEANT4~\cite{geant4} Monte-Carlo (MC) codes.
The large majority of the simulations were performed with the CYLTRAN code, mainly because this code was largely tested
for $\gamma$-ray and electron transport (e.g. in~\cite{hardy02,helmer03,helmer04}) and because it is about a factor of four
faster than the GEANT4 code. However, the CYLTRAN code has the draw-back that the geometry has to be cylindrical
which limits the flexibility of the simulations to some extent. In addition, the original version of the code did not
allow for the simulation of complete decay schemes. In fact, only a single-energy $\gamma$ ray or electron was admitted
as the input particle or an electron spectrum defined by the user. Finally, as mentioned above, positron annihilation 
in-flight was not taken into account in the code. However, it should be underlined also that the decay schemes implemented
in GEANT4 are not usable for our purpose, because first of all some of them are faulty and none of them include angular 
correlation in $\gamma-\gamma$ cascades. In the simulations, we performed, the same external event generator was used for both MC codes.

\subsection{CYLTRAN upgrade and event generation}

\subsubsection{Annihilation in-flight}

As mentioned above and found in ref.~\cite{helmer04}, annihilation-in-flight (AIF) of positrons was not included in the Monte-Carlo 
code CYLTRAN. Therefore, in the work of Helmer et al.~\cite{helmer04}, the escape probabilities had to be corrected for with a procedure 
described in the Table of Isotopes~\cite{firestone}. We corrected the CYLTRAN code and included AIF in a way consistent with all 
other interactions. The cross-section calculations
were taken from the GEANT3 code~\cite{geant3} and the probability of annihilation in-flight was added to the other interaction 
probabilties in the code where appropriate. The correct functioning of this new part of the code was tested with 
the comparison of the single- and double-escape probabilities for high-energy $\gamma$ rays interacting in the 
detector.

\begin{figure}[hht]
\begin{center}
\includegraphics[width=0.35\textwidth,angle=-90]{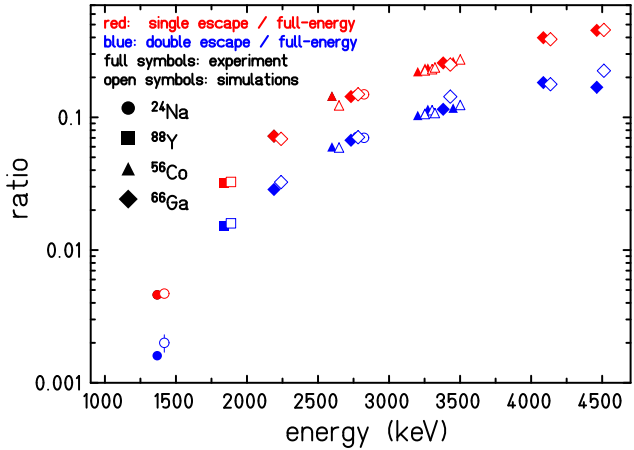}
\caption{Ratios of single-escape over full-energy intensity (red) as well as of double-escape over full-energy intensity (blue) 
         from source measurements with $^{24}$Na, $^{88}$Y, $^{56}$Co, and $^{66}$Ga. Full symbols represent experimental ratios, 
         open symbols are from simulations with the CYLTRAN code (simulation results are slightly shifted with respect to 
         experimental data for a better lisibility). Very good agreement is obtained from the lowest to the highest 
         energies.
        } 
\label{fig:escape}
\end{center}
\end{figure}

In a first step, we compared in a calculation the number of counts in the full-energy peak as well as the single- and double-escape peaks to measurements and 
found excellent agreement. A more quantitative agreement can be seen in Fig.~\ref{fig:escape}. The ratios of single-escape over 
full-energy peaks as well as of double-escape over full-energy peaks from our source measurements are compared to simulations 
with the modified CYLTRAN code. Although for some of the ratios some disagreement exists, mainly due to difficulties to fit 
relatively small peaks, a very satisfactory overall agreement is found and demonstrates that the AIF probabilities are now 
correctly taken into account in the program.

\subsubsection{Simulation of full decay events}

In order to allow for the simulation of complete decay schemes including $\beta$ decay, electron capture, $\gamma$-ray 
cascades, conversion electrons, Auger electrons, and X-rays, the code was upgraded to allow for several particles or
$\gamma$ rays as input for one event. In fact, several particles in a single event were already treated in the code, 
as long as they were produced as secondary particles during the event (e.g. by electron-positron pair production, 
by Compton scattering where an electron and a $\gamma$ ray are created). However, only one particle was allowed at 
the beginning of an event. This upgrade was done similar to the treatment of secondary particles, i.e. they were added 
in a stack of particles for an event which are then treated one by one.

The generation of a decay event is done in a separate program which generates these complete events and writes them
into a file which is then read in by either CYLTRAN or GEANT4 for the simulation of a germanium spectrum of e.g. a 
$^{60}$Co source. The input data needed to generate the decay events were principally (see Tab.~\ref{tab:sources}) taken 
from the Evaluated Nuclear Structure Data Files (ENSDF) from the National Nuclear Data Center in Brookhaven~\cite{ensdf}, 
from the Atomic Mass Evaluation~\cite{ame2012}, and from the Table of Isotopes~\cite{firestone}, appendices F-3 and F-4. 
With these inputs we were able to achieve the $\gamma$ and X-ray branching ratios given in the latest IAEA evaluations~\cite{iaea07,iaea91}

Evidently, once the input event set is correctly generated, the simulation deals correctly with effects like coincident
summing, the effect of angular correlations in $\gamma-\gamma$ cascades etc. Therefore, the yields obtained for the 
full-energy peaks observed in the simulations are directly comparable to the results from the measurements with radioactive
sources, once effects of dead-time and pile-up are corrected for or negligible. However, this procedure has also some draw-backs. 
For example, the emission angle of $\gamma$ rays in the simulations can no longer be
limited to the opening cone of the germanium detector, but a 4$\pi$ simulation has to be performed in all cases which leads
to very long simulation times. In a similar way, the environment around the detector in addition to the detector housing 
has to be present in the simulations, because the total-to-peak ratio has to be correct from the beginning and can not
be corrected by a factor as done e.g. in~\cite{hardy02,helmer03,helmer04}. However, we have now the possibility to calculate
efficiencies for single-energy $\gamma$ rays as well as for spectra from any source, as long as the necessary input data
are available.

\begin{table*}[hht]
\renewcommand{\arraystretch}{1.25}
\caption{Source characteristics used to determine the detector efficiency. We give the error bar for the half-life of $^{60}$Co, 
as we use the absolute activity of this source.       
         }                                                                        
\begin{center}
\begin{tabular}{clrrc|clrrc}
\multicolumn{10}{c}{} \\
\hline \rule{0pt}{1.3em}
~~~~Nuclide~~~~& T$_{1/2}$~~~~   &~~~~ $E_\gamma$ (keV) & ~~~~$P_\gamma$ (\%) &~~~~References~~~~  &~~~~ Nuclide~~~~& T$_{1/2}$~~~~  & ~~~~$E_\gamma$ (keV) & ~~~~$P_\gamma$ (\%) &~~~~References~~~~ \\
[0.5em] \hline \rule{0pt}{1.3em}
$^{24}$Na  &  14.9590 h      & 1368.6   &  99.9935(5) & \cite{iaea07}             & $^{133}$Ba & 11.551 y        &30.6-30.9 & 96.8(11)    & \cite{iaea91,iaea07}      \\ 
           &                 & 2754.0   &  99.872(8)  &                           &            &                 &34.9-36.0 & 22.8(3)     &                           \\ 
$^{27}$Mg  &  9.458 min      &  170.7   &   0.85(3)   & \cite{shibata98}          &            &                 &   53.2   & 2.14(3)     &                           \\ 
           &                 &  843.8   &  72.1(3)    &                           &            &                 & 79.6-81.0& 35.55(35)   &                           \\  
           &                 & 1014.4   &  27.9(3)    &                           &            &                 &   223.2  & 0.453(4)    &                           \\                                     
$^{48}$Cr  &  21.56 h        &  112.3   &  98.34(4)   &                           &            &                 &   276.4  & 7.16(5)     &                           \\
           &                 &  308.2   &  99.473(5)  &                           &            &                 &   302.9  & 18.34(13)   &                           \\
$^{56}$Co  &  77.236 d       &  846.8   & ~99.9399(23)& \cite{iaea07}             &            &                 &   356.0  & 62.05(19)   &                           \\
           &                 & 1037.8   &  14.03(5)   &                           &            &                 &   383.8  & 8.94(6)     &                           \\
           &                 & 1175.1   &  2.249(9)   &                           & $^{134}$Cs & 753.5 d         &   475.4  & 1.49(2)     & \cite{iaea91,iaea07}      \\
           &                 & 1238.3   &  66.41(16)  &                           &            &                 &   563.2  & 8.37(3)     &                           \\
           &                 & 1360.2   &  4.280(13)  &                           &            &                 &   569.3  & 15.38(4)    &                           \\
           &                 & 1771.3   &  15.45(4)   &                           &            &                 &   604.7  & 97.65(2)    &                           \\  
           &                 & 2015.2   &  3.017(14)  &                           &            &                 &   795.8  & 85.5(3)     &                           \\
           &                 & 2034.8   &  7.741(13)  &                           &            &                 &   801.9  & 8.70(3)     &                           \\
           &                 & 2598.8   &  16.96(4)   &                           &            &                 &  1038.6  & 0.990(5)    &                           \\
           &                 & 3201.9   &  3.203(13)  &                           &            &                 &  1168.0  & 1.792(7)    &                           \\
           &                 & 3253.6   &  7.87(3)    &                           &            &                 &  1365.2  & 3.017(12)   &                           \\
           &                 & 3273.0   &  1.855(9)   &                           & $^{137}$Cs &   30.09 y       &31.8-32.2 & 55.4(8)     & \cite{iaea07}             \\
           &                 & 3451.1   &  0.942(6)   &                           &            &                 &36.3-37.4 & 13.21(22)   &                           \\
           &                 & 3548.3   &  0.178(9)   &                           &            &                 &   661.7  & 84.99(20)   &                           \\
$^{60}$Co  &  1925.23(27) d  & 1173.2   &  99.85(3)   & \cite{iaea07}             & $^{152}$Eu &   13.53 y       &39.5-46.8 & 73.32(35)   & \cite{iaea07}             \\  
           &                 & 1332.5   &  99.9826(6) &                           &            &                 &   121.8  & 28.41(13)   &                           \\
$^{66}$Ga  &  9.304 h        &  833.5   &   5.906(19) & \cite{severin10,baglin02} &            &                 &   244.7  & 7.55(4)     &                           \\  
           &                 & 1039.2   &  37.07(11)  &                           &            &                 &   344.3  & 26.58(12)   &                           \\  
           &                 & 1333.1   &   1.177(4)  &                           &            &                 &   411.1  & 2.237(10)   &                           \\
           &                 & 2189.6   &   5.3457(19)&                           &            &                 &   444.0  & 3.125(14)   &                           \\  
           &                 & 2751.8   &  22.74(9)   &                           &            &                 &   778.9  & 12.96(6)    &                           \\
           &                 & 3228.8   &   1.513(7)  &                           &            &                 &   867.4  & 4.241(23)   &                           \\
           &                 & 3380.9   &   1.468(7)  &                           &            &                 &   964.1  & 14.62(6)    &                           \\
           &                 & 3422.0   &   0.858(5)  &                           &            &                 &  1085.8  & 10.13(6)    &                           \\
           &                 & 4085.9   &   1.277(7)  &                           &            &                 &  1089.7  & 1.731(10)   &                           \\
           &                 & 4806.0   &   1.865(11) &                           &            &                 &  1112.1  & 13.40(6)    &                           \\
$^{75}$Se  &  119.778 d      &   66.0   &  1.112(12)  & \cite{iaea07}             &            &                 &  1212.9  & 1.415(9)    &                           \\
           &                 &   97.0   &  3.42(3)    &                           &            &                 &  1299.1  & 1.632(9)    &                           \\
           &                 &  121.0   &  17.2(3)    &                           &            &                 &  1408.0  & 20.85(9)    &                           \\
           &                 &  136.0   &  58.2(7)    &                           & $^{180}$Hf$^m$ &  5.47 h     &    93.3  & 17.13(28)   & \cite{ensdf}              \\
           &                 &  199.0   &  1.48(4)    &                           &            &                 &   215.3  & 81.30(66)   &                           \\
           &                 &  265.0   &  58.9(3)    &                           &            &                 &   332.3  & 94.10(75)   &                           \\
           &                 &  279.0   &  24.99(13)  &                           &            &                 &   443.1  & 81.87(94)   &                           \\  
           &                 &  303.0   &  13.16(8)   &                           &            &                 &   500.7  & 14.30(28)   &                           \\
           &                 &  401.0   &  11.47(9)   &                           & $^{207}$Bi &   32.31 y       &    72.8  & 21.69(24)   & \cite{iaea07}             \\  
$^{88}$Y   &  106.625 d      &   66.0   &  1.112(12)  & \cite{iaea07}             &            &                 &    75.0  & 36.50(40)   &                           \\
           &                 &  898.0   & 93.90(23)   &                           &            &                 &    84.8  & 12.46(23)   &                           \\
           &                 & 1836.0   & 99.38(3)    &                           &            &                 &    87.6  &  3.76(10)   &                           \\  
           &                 &          &             &                           &            &                 &   569.7  & 97.76(3)    &                           \\
           &                 &          &             &                           &            &                 &  1063.7  & 74.58(49)   &                           \\  
           &                 &          &             &                           &            &                 &  1770.2  &  6.87(3)    &                           \\
[0.5em] \hline                                                                    
\end{tabular}                                                                     
\label{tab:sources}                                                               
\end{center}                                                                      
\end{table*}

\subsection{Detector model}                                                       
                                                                                  
\begin{figure}[hht]
\begin{center}
\includegraphics[width=0.24\textwidth,angle=-90]{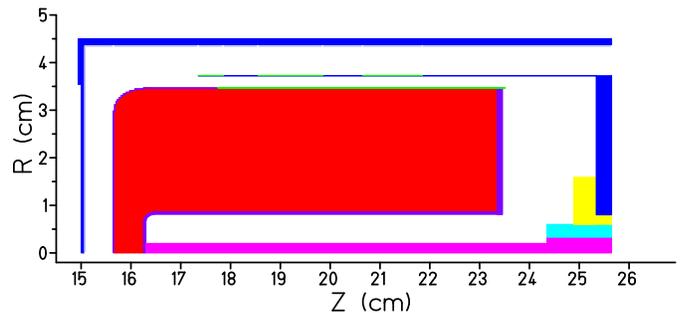}
\caption{Detector model as used in CYLTRAN simulations. The colour code is as follows: red: germanium; violet: germanium dead material; 
         blue: aluminum; pink: brass. The green material corresponds to screws from the detector holding structure, which, in CYLTRAN, 
	     had to be "distributed" over a hollow cylinder. The other colours correspond to isolating material and teflon.  
         } 
\label{fig:model}
\end{center}
\end{figure}

The detector model used in the simulations is shown in Fig.~\ref{fig:model}. The version shown is the one used in the
CYLTRAN simulations. However, the only differences between the CYLTRAN and GEANT4 models come from the fact the CYLTRAN 
knows only cylinders. Therefore, the round edges of the germanium crystal had to be constructed by adding many tiny cylinders
to achieve an apparent round edge by keeping the germanium volume constant. In GEANT4, the use of a torus greatly 
\begin{figure*}[hht]
\begin{center}
\includegraphics[width=0.43\textwidth,angle=-90]{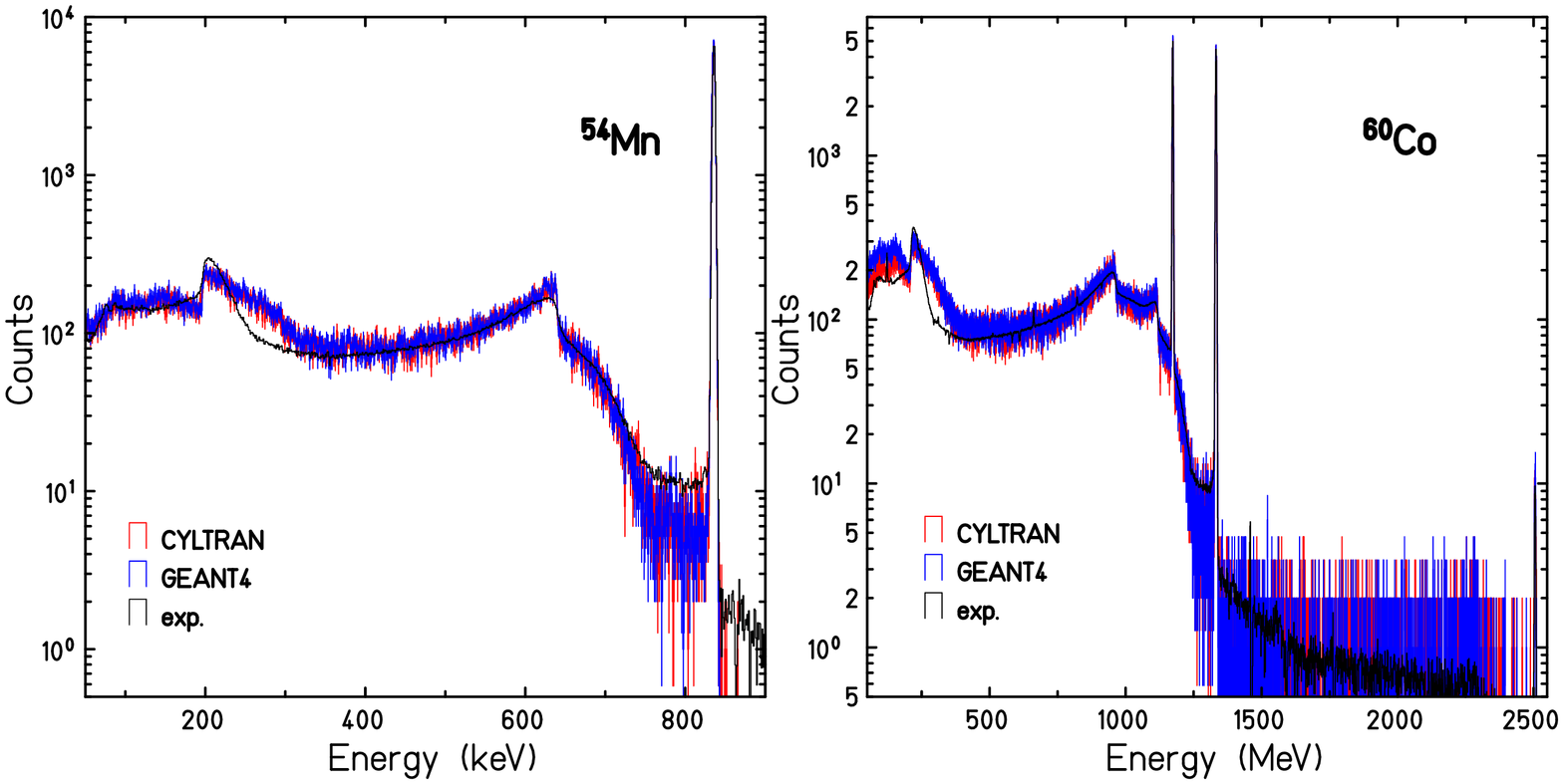}
\caption{Comparison of CYLTRAN and GEANT4 simulation with measurements using sources of $^{54}$Mn and $^{60}$Co with the final 
         detector model including the environment around the detector.
         The different spectra are matched with the height of the full-energy peaks. 
         } 
\label{fig:mn54_co60}
\end{center}
\end{figure*}
simplifies the modelling task. The dimensions used in the final detector model are compared to the manufacturer data in
Tab.~\ref{tab:det}. It is interesting to note that the central hole of the crystal is much longer and wider than given
originally by the vendor.

Fig.~\ref{fig:mn54_co60} compares the simulation results from CYLTRAN and GEANT4 to experimental spectra. Although some 
deficiencies can be observed, e.g. a lack of counts below the full-energy peak and an overestimation of the number of counts
and a different shape around the edge from 180$^o$ backscattering of Compton scattering events around 200-300~keV, 
the overall shape agreement is quite satisfying.

In these simulations, we used a detector resolution as determined experimentally. However, in the simulations which we discuss in the
following and which were used to determine the peak and total efficiencies, we did not use a detector resolution in order to more
easily determine the number of counts in a peak. The number of counts was determined from the number of counts in the channel
of the full-energy peak from which the average number of counts of the five neighbouring channels left and right from the peak
were subtracted as the background under the peak.

\section{Analysis of experimental data}

\subsection{Analysis of detector scan data}

\begin{figure}[hht]
\begin{center}
\includegraphics[width=0.48\textwidth,angle=0]{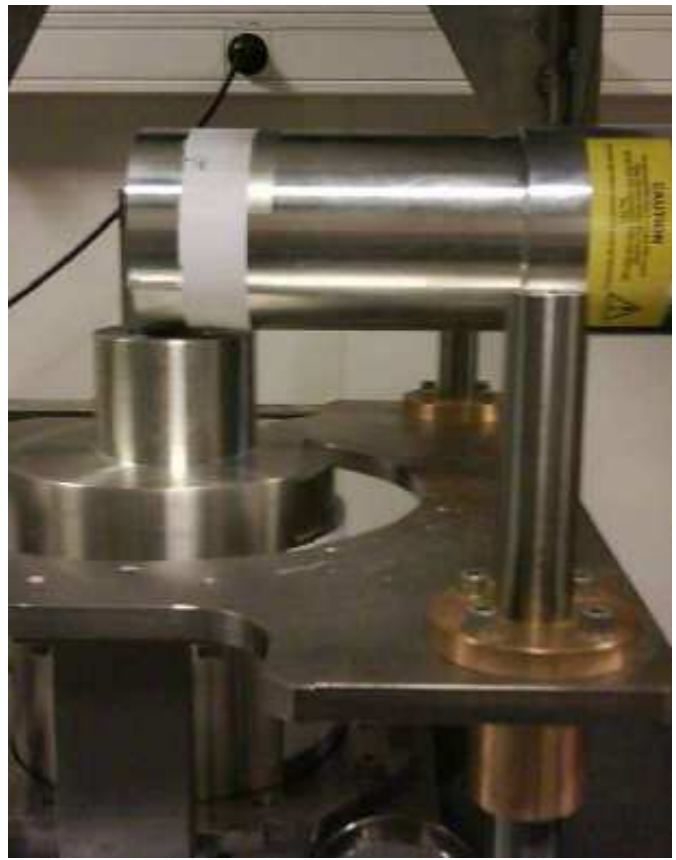}
\caption{Photograph of the germanium detector installed on the scanning table at CSNSM. In the present position, 
         the crystal side was scanned in two directions. 
         } 
\label{fig:agata}
\end{center}
\end{figure}

The detector scans were performed at the AGATA detector scanning table at CSNSM~\cite{ha13} and at CENBG. 
At CSNSM, a strongly collimated $^{137}$Cs source with a $\gamma$-ray energy of 662~keV and an activity of about 
477~MBq was used. Fig.~\ref{fig:agata} shows the detector installed on the X-Y scanning table at CSNSM.
One- and two-dimensional scans were performed over the front of the detector as well as over the full length of the 
crystal. Typical step sizes were between 0.5~mm (one-dimensional scans) and 2~mm (two-dimensional scans). 
A measurement point lasted typically 20 s. The data were analysed by either integrating the full-energy peak 
or the total number of counts in the spectrum thus neglecting the small contribution from room background. 
As will be shown below, these scans allowed to fine tune the parameters of the detector model.

At CENBG, scans were performed with low-energy sources of $^{57}$Co and $^{241}$Am. In these scans, the sources
were collimated only in one dimension, i.e. the sources viewed the detector through a one-millimeter slit. Steps of 1~mm
and measurement times of 30~min were used. In this case, only the full-energy peak was integrated as a function of the
position of the source.

\begin{figure*}[hht]
\begin{center}
\includegraphics[width=0.43\textwidth,angle=0]{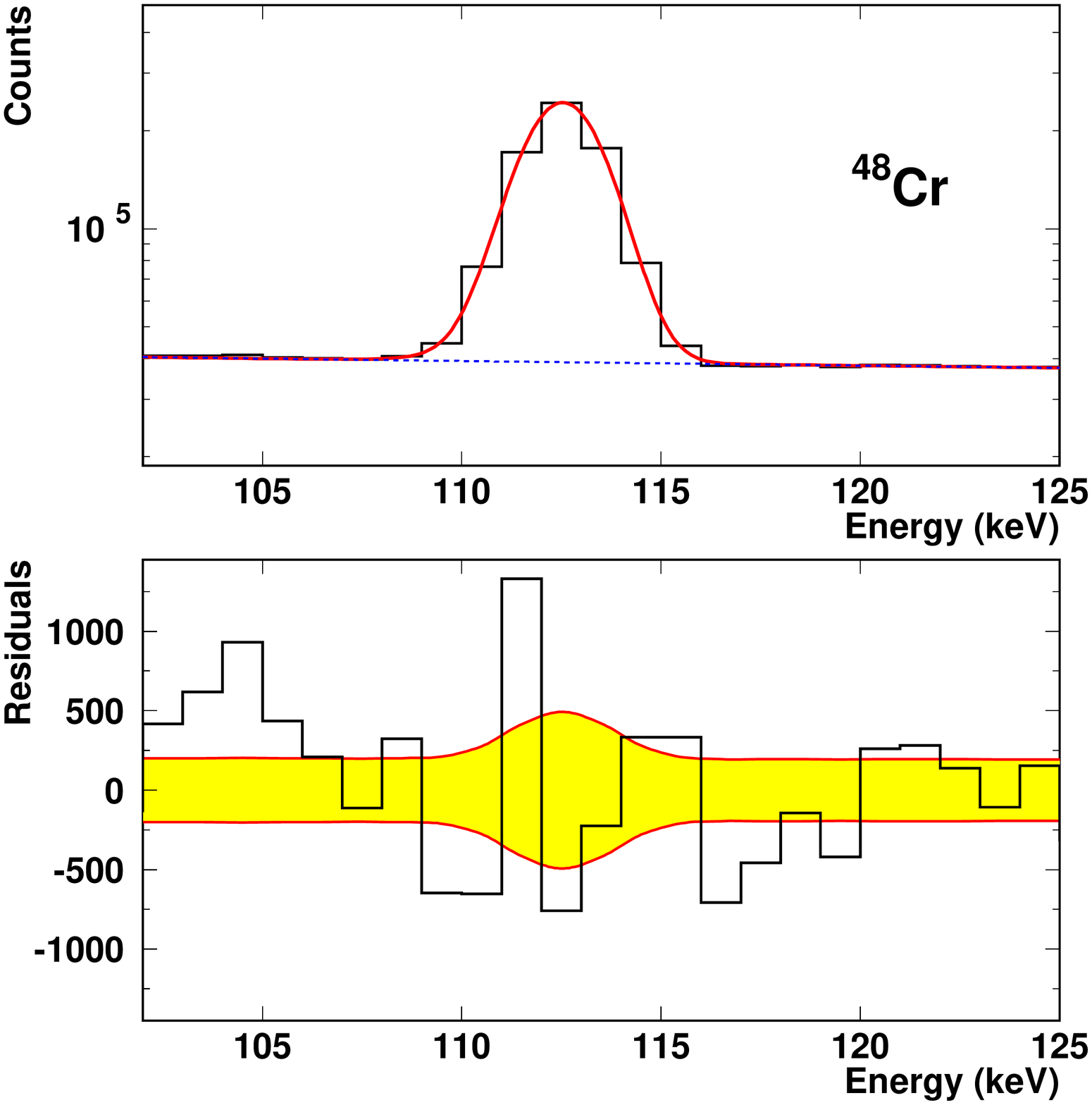} 
\includegraphics[width=0.43\textwidth,angle=0]{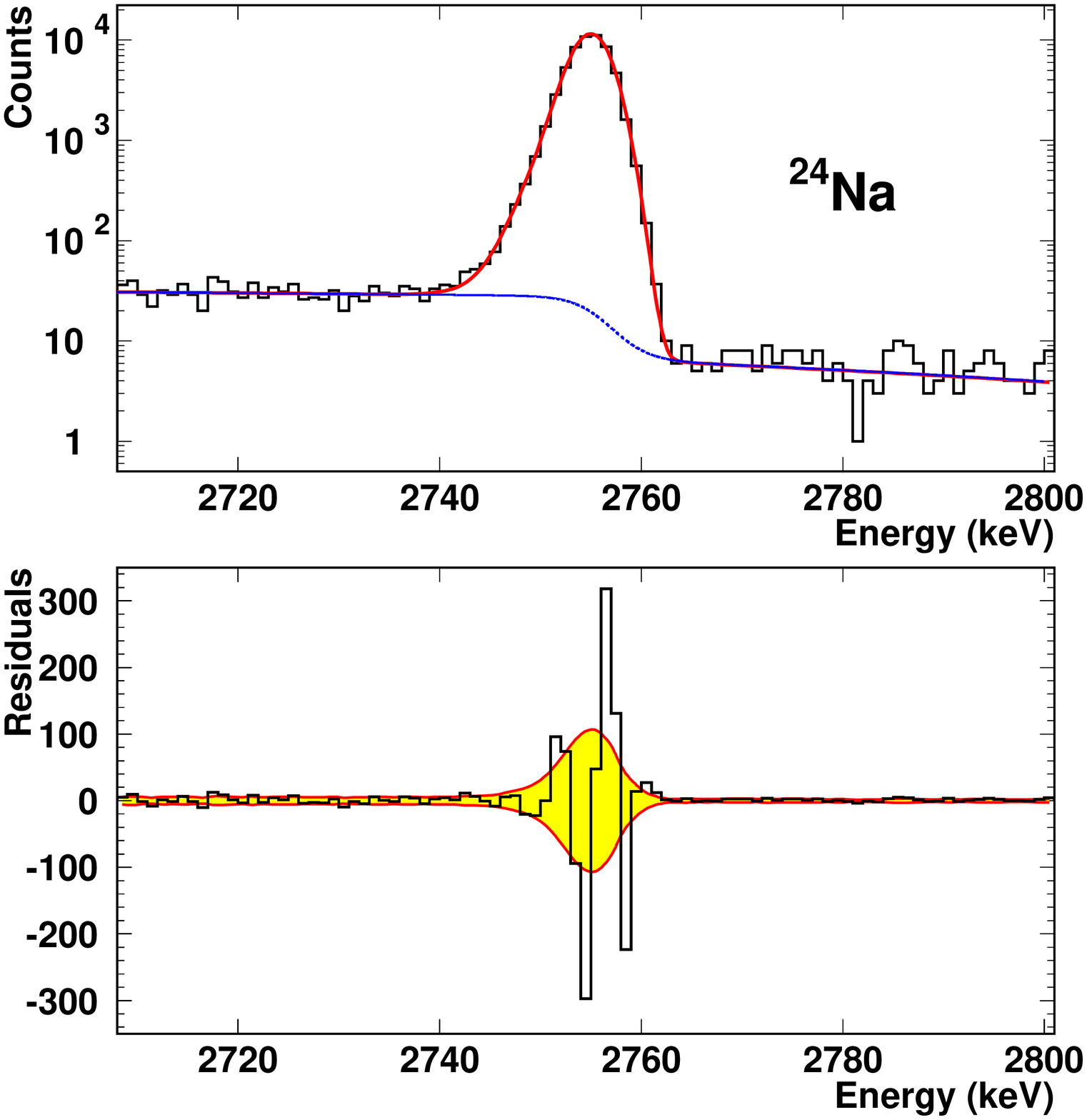}
\caption{Comparison of an experimental spectrum for a low-energy and a high-energy peak with the fit. 
         The left-hand side shows a fit of the 112~keV peak of $^{48}$Cr and the right-hand side a fit of the 2754~keV line
         of $^{24}$Na. The histogram is the experimental spectrum, the red line is the fit, and the dashed blue line is the 
	 background function. The low-energy peak is fitted with a second-degree polynomial as the background, 
         whereas a smoothed step function is used for the high-energy peak. 
         The lower figures show the fit residuals. The black histograms are the residuals and the yellow bands are the statistical 
         error bands.
         } 
\label{fig:peaks}
\end{center}
\end{figure*}

\subsection{Analysis of the radioactive source data}

The experimental data for all source measurements were taken with a GANIL-type data acquisition~\cite{ganildaq}. The data were analysed 
using the PAW software~\cite{paw}. In contrast to the work of Helmer {\it et al.}~\cite{helmer03} who used only a simple Gaussian to fit 
the full-energy peaks, we fitted them with a function containing a Gaussian plus
a low-energy tail described by a shifted asymmetric Gaussian to take into account the incomplete charge collection 
in the germanium crystal. To constrain the fit and reach a faster convergence, the tail parameters were first determined
over a wide range of energies and then fixed in the fits as a function of the peak energy.

As the background function, we used three different descriptions: (i) a second-order polynomial, (ii) a linear step function
(complement of an error function and a straight line), or (iii) two independent linear functions left and right of the peak "smoothly"
connected under the peak. Out of these three functions, we chose for each peak the two most appropriate background functions 
to fit the peaks. For each peak, two fits were performed over a different fitting window yielding thus four different fits
for each peak. The best fit in terms of the normalised $\chi^2$ was kept with its error as determined from the fit.
The second best fit was used to determine a systematic error from the difference of the two best fits which was added quadratically
to the fit error from the best fit.

Fig.~\ref{fig:peaks} shows the fit result for a low-energy and a high-energy peak together with the residuals from
the experimental spectrum and the fit. Although it is evident that the function used to describe the full-energy peak
is not complete, the integral of the peak is determined with high precision.

\begin{figure*}[hht]
\begin{center}
\includegraphics[width=0.5\textwidth,angle=-90]{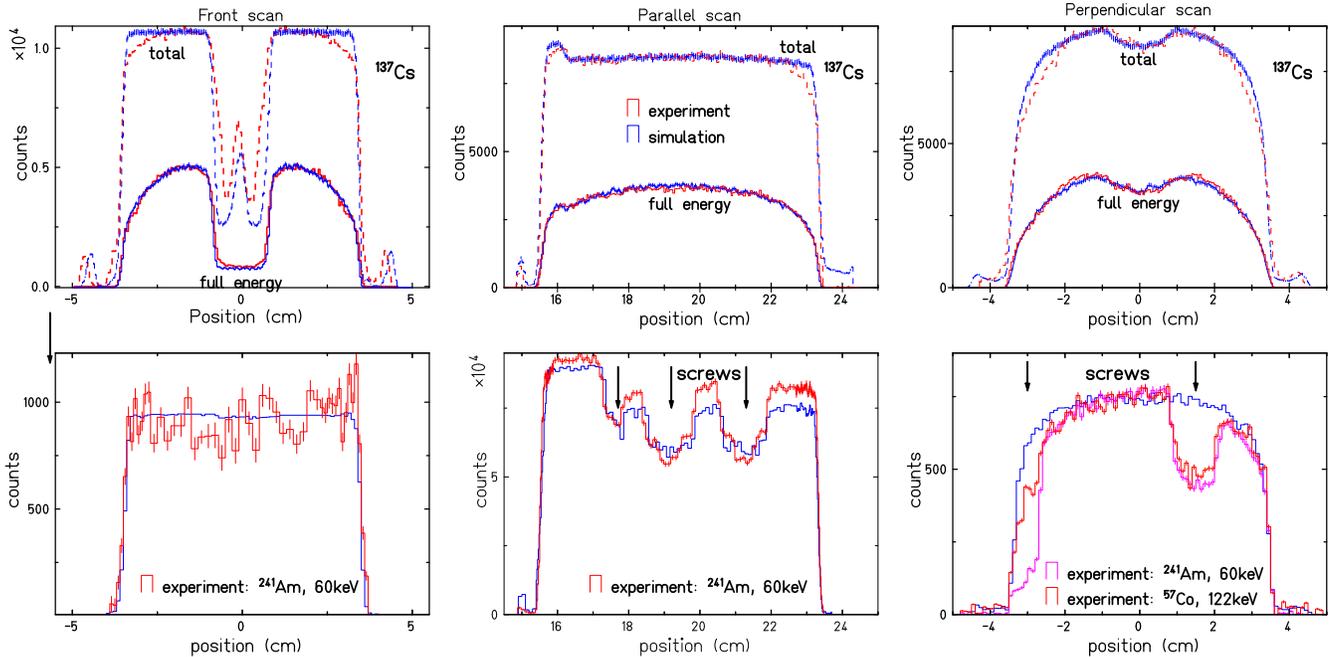}
\caption{Comparison of experimental data from the scans of the different detector dimensions with simulation performed
         with the CYLTRAN code. From left to right: Front scan, parallel scan, perpendicular scan. The $^{137}$Cs scans 
         (top panels) yield a clear picture of the size of the crystal and its position in the detector can. In particular, 
         the tilt of the crystal in the aluminum can can be seen by means of a shift of Compton scattering events from the
         detector can and the central rod between simulation and experiment, while the germanium crystal is perfectly aligned.
         In the total energy spectrum, the distance of the crystal from the entrance window could be determined in the parallel
         scan. The low-energy scans (lower panels) are affected by the holding structure of the crystal, mainly by the screws 
         indicated in the figure, which absorbs to some extent these low-energy $\gamma$ rays. This is seen with the scans with
         the $^{57}$Co source (E$_\gamma$ = 122~keV) and the $^{241}$Am source (E$_\gamma$ = 60~keV). In the perpendicular scan, 
	 the screws are not seen in the simulation, as they are "distributed" over a cylindrical ring.
         } 
\label{fig:scans}
\end{center}
\end{figure*}

\section{Experimental results and comparison with CYLTRAN simulations}

In the following sections, we will first present the results from the detector scans  which allowed for a fine
adjustment of the detector model parameters. With this detector model, we will then compare the source measurements
to simulations in order to verify that the detector model allows for an overall agreement 
between simulation and measurement over a wide range of $\gamma$-ray energies.

\subsection{Results from detector scans}

We have scanned the detector in three direction: (i) a scan on the front side of the detector, (ii) a scan parallel to the 
cylinder axis of the crystal, and (iii) a scan perpendicular to the cylinder axis. They are termed "front scan", "parallel scan", 
and "perpendicular scan", respectively, in the following.

Fig.~\ref{fig:scans} shows a comparison of the experimental scan results and the simulations. In particular, the scans with
the $^{137}$Cs source allowed us to precisely adjust the detector model parameters. The figure shows that nice
agreement is obtained for the full-energy peak for all scans, whereas some deficiencies remain for the total number of
counts observed. This is due to the fact that for the full energy peak intensity, only the detector crystal itself is 
important, while for the total spectrum the environment around the detector influences the counting rate. As this environment
is different from one experimental site to another, we did not insist too much in improving this aspect. The environment
was adjusted in the simulations in order to reproduce the Total-to-Peak ratio (see below), the only important parameter
related to the physical environment around the detector.

The scans with the low-energies sources performed at CENBG did not contribute as much to the detector parameter determination.
This is mainly due to the fact that low-energy $\gamma$ rays are, at least to some extent, absorbed in the material around
the detector and notably by the crystal holding structure, as can be seen by the scans with the $^{57}$Co and 
$^{241}$Am sources. The screws used to fix the crystal within its housing absorb clearly a large part of the $\gamma$ rays.
Nonetheless, these measurements allowed us to verify the parameters determined with the high-energy scan.

On the parallel scan with the $^{137}$Cs source, one can see that the height of the total spectrum in the simulation is
slightly larger than in the experimental data in the region of the entrance window. This might be an indication that 
the entrance window of the aluminum can is thinner than given by the manufacturer. Evidently this would influence 
the efficiency in particular for low-energy $\gamma$ rays. In order to check this thickness, we performed measurements 
with a mono-energetic electron source.

\subsection{Thickness of detector entrance window}

At CENBG, the "Neutrino" group possesses a mono-energetic electron source made of a collimated electron beam from a strong 
$^{90}$Sr source which passes through a magnetic field to select a particular energy bin. After an energy calibration with 
low-energy $\gamma$-ray sources, we used electron energies of 0.8, 1.0, 1.4, and 1.8~MeV incident on the entrance window
in a direction parallel to the detector axis. The electron peak position and thus the energy loss in the entrance window
were compared to simulation with the CYLTRAN code for the same conditions. 

\begin{figure}[hht]
\begin{center}
\includegraphics[width=0.5\textwidth,angle=0]{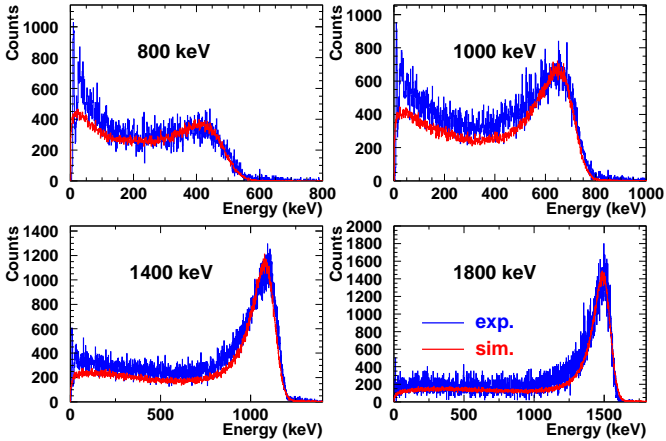}
\caption{Comparison of experimental data and simulations for an electron impact perpendicular to the detector entrance 
          window at different energies. The simulations are performed with an entrance window thickness of 0.7~mm of aluminum, 
          the value given by the manufacturer.
         } 
\label{fig:electrons}
\end{center}
\end{figure}

In the simulations, we varied the thickness of the entrance window. As can be seen from Fig.~\ref{fig:electrons}, it is 
somewhat difficult to determine the exact peak position of the electrons, as the peak is much larger than a typical 
$\gamma$-ray peak. However, within a thickness range of 0.65 and 0.7~mm we found good agreement between experimental 
measurements and simulations. This thickness range includes the manufacturer's value of 0.7~mm. In the following, we
will therefore use a thickness of 0.7~mm.

\subsection{Results with $\gamma$-ray sources}

In the following paragraphs, we will present and discuss the results obtained by the measurements with the different
radioactive sources. The Total-to-Peak ratio has been determined with ten different sources, whereas we used 14 different
radioactive sources for the relative full-energy efficiency. These 14 sources are presented in Tab.~\ref{tab:sources}.
As standard calibration sources do not cover the full range of energy of interest in the present work, we produced also 
short-lived sources to cover the full energy range. None of the sources, except $^{60}$Co, had activities which were 
known to a precision at the per mil level. This is of course particularly true for the short-lived sources
produced online. Therefore, only two $^{60}$Co sources with an activity known to 0.09\% were used for an absolute calibration
of the efficiency curve. A second method to determine the absolute efficiency, again for the $^{60}$Co $\gamma$ rays,
used the coincident summing of the two $\gamma$ rays.

For all standard sources, the activity is deposited on a thin plastic foil and covered with another thin foil which
does not reduce in any significant way the $\gamma$-ray counting rate. These foils are held by a plastic frame with a diameter
of 25~mm or 37~mm and thicknesses of typically 1-2~mm.

Most of the short-lived sources were produced at ISOLDE by depositing the activity, after online mass separation, on a
plastic catcher of 25~mm diameter and a thickness of 4~mm. With a total energy of 40~keV, the isotopes are implanted
at the surface and thus again the $\gamma$-ray absorption in negligible. Sources of $^{24}$Na, $^{27}$Mg, $^{41}$Ar,
$^{48}$Cr, $^{56}$Co, $^{58}$Co, $^{65}$Zn, $^{66}$Ga, $^{75}$Se, and $^{180}$Hf$^m$ were thus produced. 

In addition, stronger sources of $^{56}$Co and $^{66}$Ga were produced by (p,n) reactions at the Tandem of IPN Orsay.
In this case, the activity was produced over the whole target thickness (5$\mu$m of zinc for the $^{66}$Ga activity, 
56$\mu$m of $^{56}$Fe for the $^{56}$Co source). However, dealing only with rather high $\gamma$-ray energies for 
these two sources, there was also no sensible attenuation of the $\gamma$ rays.

\subsubsection{Total-to-peak ratio}

To measure the Total-to-Peak (T/P) ratio, the ideal source has one $\gamma$ ray with 100\% branching ratio and no 
other radiation. However, such a source does not exist. From the sources used, $^{54}$Mn comes closest to this ideal 
source. The sources chosen for the T/P ratio measurement come relatively close to this ideal case, some of them emitting additional
X rays, low-energy $\beta$ particles, or 511~keV annihilation radiation. The $^{60}$Co source has two relatively close 
lying $\gamma$ rays which can be treated as one single energy.

The sources used are the following: $^{57}$Co, $^{51}$Cr, $^{85}$Sr, $^{137}$Cs, $^{58}$Co, $^{54}$Mn, $^{65}$Zn, 
$^{60}$Co, $^{22}$Na, and $^{41}$Ar. Source measurements were combined with background measurements for each source.

\begin{figure}[hht]
\begin{center}
\includegraphics[width=0.38\textwidth,angle=-90]{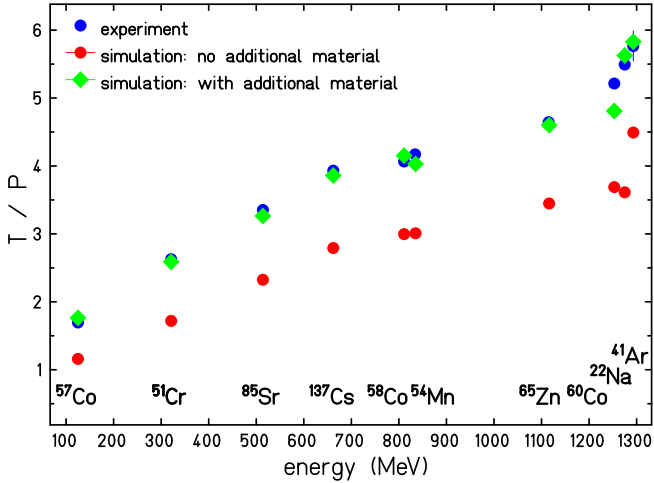}
\caption{Total-to-Peak ratio for $\gamma$-ray energies ranging from 122~keV to 1300~keV. As the measurements as well as the simulations
         are performed with complete decay schemes, the T/P ratio does not grow smoothly with energy. The details depend
         rather on the presence of other radiation. What is important is the overall good agreement between experiment
         and simulation.
         } 
\label{fig:t-to-p}
\end{center}
\end{figure}

Fig.~\ref{fig:t-to-p} shows the experimental results. These measurements are compared first with simulations where
only the germanium detector was modeled, but where the "environment" was not taken into account. Evidently, the simulated T/P 
ratio is significantly lower than the experimental one. Only when surrounding materials such as the measurement table, the source holder, 
the walls etc. are taken into account, an overall very good agreement could be obtained. The difficulty is to include
this external material in CYLTRAN simulations. We achieved the good agreement shown in the figure by adding a hollow aluminum
cylinder of thickness 1.6~mm and inner diameter 6.7~cm and an aluminum disk of 4~cm thickness and of radius 4.5~cm
behind the source. As the detailed geometry of the backscattering material is not of interest (except if we would
like to improve on the agreement between experiment and simulation of Figs.~\ref{fig:mn54_co60}), we kept this backscattering
configuration.

\subsubsection{Absolute efficiency with $^{60}$Co}

The absolute efficiency for the two $\gamma$-ray energies of $^{60}$Co, i.e. 1173~keV and 1332~keV, was determined
in two independent ways and with different sources. The first determination was done with two $^{60}$Co sources calibrated 
in activity with a precision of 0.09\%. The second determination was done using the sum-energy peak from the coincident
summing of the two $^{60}$Co $\gamma$ lines.

\paragraph{Absolute efficiency with activity calibrated $^{60}$Co sources}

The procedure used in this paragraph is the standard procedure to determine the absolute efficiency of $\gamma$-ray detectors:
knowing the activity of the source ($A$), the measurement time ($t$), and the branching ratios ($BR$) of the $\gamma$ rays 
emitted, and by determining the number of counts ($N$) in the full-energy peak, the efficiency $\epsilon^{FE}$ is determined 
as follows:
$$
\epsilon^{FE} = \frac{N}{A \times t \times BR}.
$$

The number of counts in the full-energy peak has to be corrected for the acquisition dead-time and pile-up effects
as described above. In principle, the coincident summing effect has to be taken into account as well. However, as we only 
compare source data to simulations with the full decay scheme of a source, these summing effects are present in both
data, the experimental data and the simulations.

The main problem in this procedure is to have a source with an activity precision of the order of 0.1\%. We possess two different
sources with this precision prepared by the LNHB laboratory at Saclay, France~\cite{lnhb}. They have an absolute uncertainty on
their activity of 0.09\%. These sources allow us to determine the efficiency of our detector at the $^{60}$Co $\gamma$-ray energies
of 1173~keV and 1332~keV. These efficiencies are 0.2175(3)\% at 1173~keV and 0.1996(3)\% at 1332~keV.

These measurements have been performed at different periods over more than one year. During this time, the detector was warmed up 
at several occasions, also for longer periods of several weeks. However, we did not observe any change in efficiency and conclude
thus that the efficiency is constant over time.

\paragraph{Absolute efficiency from $\gamma-\gamma$ coincidences}

The absolute efficiency of a germanium detector can also be determined from $\gamma-\gamma$ coincidences. For this purpose, 
one needs a source with a high-branching-ratio $\gamma-\gamma$ cascade without any cross-over $\gamma$ ray. The $^{60}$Co as well
as the $^{24}$Na sources have these characteristics. However, the short half-life of $^{24}$Na prevents from using this source 
for this purpose.

We have performed a series of measurements with a 25~kBq $^{60}$Co source. By determining the number of counts in the two $\gamma$-ray
peaks at 1173~keV and at 1332~keV as well as in the sum peak at 2505~keV, one can establish three equations for three unknowns:

\begin{eqnarray}
  N_{1173} & = & \epsilon^{FE}_{1173} * A * BR_{1173} * (1 - \epsilon^t_{1332} * w_{12}) \nonumber \\
  N_{1332} & = & \epsilon^{FE}_{1332} * A * BR_{1332} * (1 - \epsilon^t_{1173} * w_{12}) \nonumber \\
  N_{2505} & = & \epsilon^{FE}_{1173} * \epsilon^{FE}_{1332} * A * BR_{2505} * w_{12} \nonumber
\end{eqnarray}

\noindent
$N_x$ are the numbers of counts in the full-energy peaks at energy $x$, $A$ is the source activity, $BR_x$ is the branching ratio for 
the different energies ($BR_{2505}$ is the probability to have both $\gamma$ rays, the 1173~keV and the 1332~keV $\gamma$ rays, in 
coincidence), $\epsilon^t$ is the total efficiency, and $w_{12}$ is the correction due to the $\gamma-\gamma$ angular correlation. 

In these equations, the three unknowns are the efficiencies at the two energies and the activity. All the other quantities are either 
determined from the spectrum ($N_x$), from other measurements ($\epsilon^t$), or from calculations ($w_{12}$). In the present procedure, 
the acquisition dead-time as well as, to a large extent, pile-up can be neglected as they affect all peaks in the same way. However, 
there is one effect which has to be corrected for for the sum peak. This is the probability that a 1173~keV $\gamma$ ray from one event 
is added to a 1332~keV $\gamma$ ray from another event and vice versa and add to the sum energy peak at 2505~keV. This effect 
is not negligible, as can be seen from the presence of small but visible peaks at 2346~keV and 2664~keV, twice the energies of the 
individual $\gamma$ rays. We used the counting rates in these two sum peaks, corrected for the detection efficiency of the 
respective other $\gamma$-ray energy, and determined thus the number of counts to be subtracted from the 2505~keV sum peak.

The angular correlation correction $w_{12}$ was determined in a Monte-Carlo simulation with the CYLTRAN code, where we compared 
the sum energy peak in a simulation with angular correlation to a simulation for an isotropic emission of the $\gamma$ rays. 
The result of these simulations is quite close to a calculation of the angular correlation correction at a fixed angle of 0$^o$ 
which does not take into account the opening angle of the detector and thus the contribution of $\gamma-\gamma$ angles larger 
than zero.

From our data, we determine efficiencies of 0.2186(7)\% and 0.1996(7)\% at 1173~keV and 1332~keV, respectively. These efficiencies are in 
excellent agreement with the values determined with the high-precision sources. The precision is somewhat less than from the 
high-precision sources. This is exclusively due to the fact that one needs a high number of counts in the sum energy peak 
and this implies very long measuring times. In our case, we performed measurements over several months.

\paragraph{Final efficiency for the $^{60}$Co $\gamma$-ray energies}

From both the above methods to determine the absolute efficiency for the $^{60}$Co $\gamma$ rays, we arrive at a final efficiency 
of 0.2177(4)\% and 0.1996(3)\% at 1173~keV and 1332~keV, respectively. These experimental efficiencies can be compared to the 
results of our MC simulations of 0.2174(3)\% and 0.1997(3)\%. This precision corresponds to about 1.5-2~\permil.

\subsubsection{Full-energy peak efficiency with other sources}

The sources used for the determination of the full-energy efficiency curve are given in Tab.~\ref{tab:sources}
with the information pertinent for the present work. These sources cover an energy range from 30~keV to 3.4~MeV.
In our work we did not use the 500~keV $\gamma$ ray of $^{180}$Hf$^m$ and the two highest energy $\gamma$-ray lines of 
$^{66}$Ga. These $\gamma$ rays were already rejected in the work of Hardy and co-workers~\cite{hardy02,helmer03,helmer04}.
They have an apparent wrong branching ratio in the literature. 

As already mentioned, all sources were used to establish the relative full-energy peak efficiency, their activity being
treated as a free parameter in the fitting of the relative efficiency curve. This was necessary because none of the sources 
(except the two $^{60}$Co sources) is available with a precision on the activity necessary for the present work. However, our 
procedure allows us to determine with good accuracy the activity of these sources. A comparison of the activities determined 
and those given by the manufacturers is shown in Tab.~\ref{tab:activity}.

\begin{table}[hht]
\renewcommand{\arraystretch}{1.25}
\caption{Comparison between the activity of commercial sources as given by the manufacturers and the activity determined in our 
         fitting procedure. The factor given in the table is the ratio between the activity given by the manufacturer and 
         the one determined by our means. The error bars are 1-sigma error bars.      
         }                                                                        
\begin{center}
\begin{tabular}{c|cccccc}
\hline \rule{0pt}{1.3em}
~~source~~ & ~~$^{133}$Ba~~ & ~~$^{207}$Bi~~ & ~~$^{134}$Cs~~ & ~~$^{137}$Cs~~ & ~~$^{152}$Eu~~ & ~~$^{88}$Y~~ \\
factor &   1.034(9) & 1.012(5)  & 1.009(40)  & 1.020(11)  & 1.014(13)  & 1.069(8)   \\
[0.5em] \hline                                                                    
\end{tabular}                                                                     
\label{tab:activity}                                                               
\end{center}                                                                      
\end{table}

\begin{figure}[hht]
\begin{center}
\includegraphics[width=0.5\textwidth,angle=0]{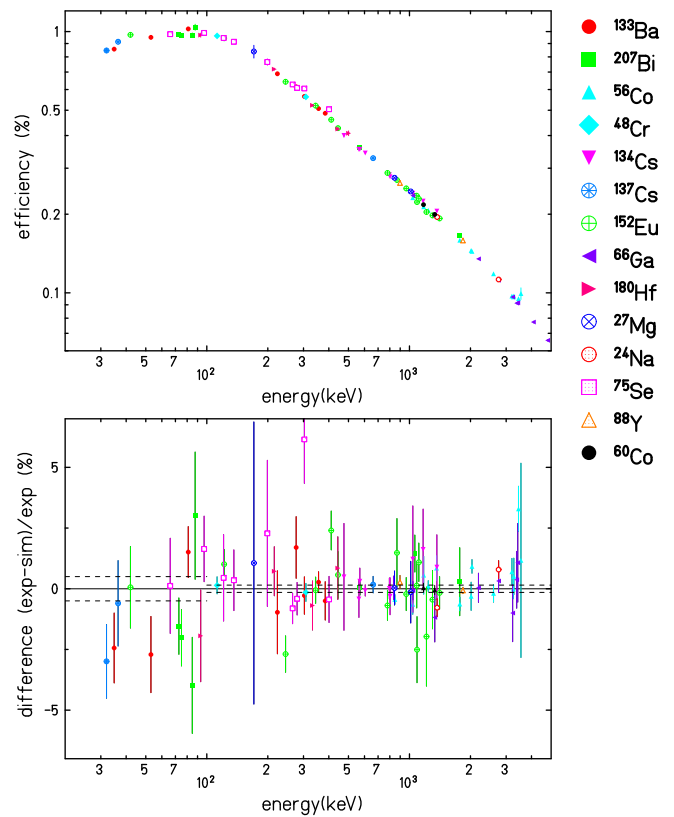}
\caption{(a) Absolute $\gamma$-ray efficiency at a distance of 15~cm between the source and the detector entrance window.
         As explained in the text, the shape of the curve is determined with the $\gamma$ rays given in 
         Tab.~\ref{tab:sources}, whereas the absolute height of the curve was determined by means of $^{60}$Co sources.
         The curve is not completely smooth, as what is presented is not the single $\gamma$-ray efficiencies, but
         full-energy peak efficiencies determined with the complete decay schemes from the sources.
         (b) Relative differences (in \%) between the experimental data and the simulations with the detector model are
         presented. The dashed lines give the final precisions adopted.
         } 
\label{fig:eff}
\end{center}
\end{figure}

Fig.~\ref{fig:eff}a shows the efficiency curve as determined from the sources with a maximum of the efficiency
of about 1\% at 60~keV (this curve is already adjusted in absolute height with the $^{60}$Co sources, see above).
A much more detailed picture is shown in Fig.~\ref{fig:eff}b, where we present the difference between the experimental
efficiency and the simulated one normalised by the experimental efficiency. Although some points lie somewhat far from
the zero-difference line, the vast majority of the data points lie within the $\pm$1\% boundaries. A fit with a constant
of these differences between experiment and simulations yields a value of (-0.002$\pm$0.061)\% with a normalised 
$\chi^2$ of 1.4. A linear fit gives a linear term of (0.16 $\pm$ 0.16) compatible with zero. The normalised $\chi^2$
does not change, another indication that the addition of a linear term does not improve the fit. If we remove, for a test, 
the data points below 50~keV, we obtain a value of (0.07 $\pm$ 0.16) for the linear term which shows that the possible
slight trend is entirely due to the low-energy points which are extremely difficult to fit in a $\gamma$-ray spectrum.

We investigated also, whether we get better agreement in the low- or high-energy region. Therefore, we fitted the difference
data in the region below 1500~keV and above 1000~keV. For the low-energy region, we get an average value for the difference
of (-0.010 $\pm$ 0.077)\%, whereas the result for the high-energy part is (-0.009 $\pm$ 0.083)\%. In both cases, a linear fit
does not improve the results.

As the result of our MC simulations exactly fits our measured efficiencies for the $^{60}$Co source and the ratio between 
experimental results and simulations are constant with energy, we conclude that we have obtained a precision on the efficiency 
simulation of our detector below 0.1\% over a wide energy range. To be conservative, we attribute an 
uncertainty of 0.5\% to our efficiency below 100~keV and of 0.1\% between 100~keV and 3.4~MeV for the relative efficiencies. 
If we combine the uncertainty of the efficiencies for the $^{60}$Co $\gamma$ rays with the uncertainty of these relative
efficiencies, we find an uncertainty of 0.5~\% below 100~keV and of 0.15~\% above. 

These values can be used now in experiments for the $\gamma$-ray efficiency precision of the detector installed at 
exactly 15~cm from the radioactive source. This is indeed the case, as the detector was used in two experiments at GANIL, 
one on LISE3 and one at the SPIRAL identification station~\cite{grinyer14}, and in an experiment at the IGISOL4 facility of 
Jyv\"a{}skyl\"a{}.

\section{Conclusions}

In the present work, we have combined source measurements and simulations to develop a germanium detector model which allows 
to describe the experimental results with simulations. For this purpose, a total of 23 different radioactive sources 
have been used. The detector model was established with scans of the germanium crystal and then compared to the source 
measurements.

This work demonstrates that we have been able to efficiency calibrate our single-crystal high-purity germanium detector with a 
precision of 0.15~\% between 100~keV and 3.4~MeV and of 0.5~\% below 100~keV. This allows us now to determine branching 
ratios of e.g. super-allowed $\beta$ decay with a similar precision.

The detector model was mainly developed with the aid of the CYLTRAN program package. However, the detector 
model was transposed recently to the GEANT4 program package and yields results in perfect agreement with the CYLTRAN simulations. 
The GEANT4 software has the advantage that more complicated experimental situations can be modelled more correctly. However, 
for the moment, the CYLTRAN description was found to be sufficiently precise for our applications.
In addition, the CYLTRAN simulations are much faster than the GEANT4 simulations.

As the detector was already used and will be used in the future in different experimental environments, each new experimental 
arrangement requires a new measurement of the total-to-peak ratio. Therefore, relatively simple sources have to be used under 
online conditions to achieve this. Typically, sources of $^{57}$Co, $^{54}$Mn, and $^{60}$Co are used for this purpose.

\section*{Acknowledgment}

We are grateful to A. Korichi and H. Ha for their help with the detector scan at CSNSM.
We thank K. Johnston, A. Gottberg, and G. Correia from ISOLDE for their help during the sample collection at ISOLDE.
Our gratitude goes also to the accelerator staff at IPN Orsay.
We are in debt to J.L. Tain for providing us with his $\gamma$-$\gamma$ correlation program and the basic version of the
event generator program. We are grateful to C. Serna for his help with the electron source.

\end{document}